\documentclass[showpacs,aps]{revtex4}
\usepackage{bm}
\usepackage{graphicx}
\usepackage{amsmath}

\def\be{\begin{equation}}
\def\lan{\left\langle}
\def\ran{\right\rangle}
\def\ee{\end{equation}}
\def\barr{\begin{array}}
\def\earr{\end{array}}

\def\nn8{\nonumber\\[15pt]}
\def\l{\left}
\def\r{\right}
\def\dis{\displaystyle}
\def\ed{\end{document}}

\def\cod{{\cal O}^\dagger}
\def\co{{\cal O}}
\def\cg{{\cal G}}
\def\ce{{\cal E}}

\begin{document}

\title{Bivariate $t$-distribution for transition matrix elements in 
Breit-Wigner to Gaussian domains of interacting particle systems} 

\author{V.K.B. Kota$^a$, N.D. Chavda$^b$ and R. Sahu$^c$}

\affiliation{$^a$Physical Research Laboratory, Ahmedabad 380 009, India \\
$^b$Department of Physics, Faculty of Science, M.S. University of Baroda, 
Vadodara 390 001, India \\ $^c$Department of Physics,
Berhampur University, Berhampur 760 007, India}

\begin{abstract}

Interacting many-particle systems with a mean-field one body part plus a
chaos generating random two-body  interaction  having strength $\lambda$,
exhibit Poisson to GOE and Breit-Wigner (BW) to Gaussian transitions in level
fluctuations and strength functions with transition points marked by
$\lambda=\lambda_c$ and $\lambda=\lambda_F$, respectively; $\lambda_F >>
\lambda_c$. For these systems theory for  matrix elements of one-body
transition operators is available, as valid in the Gaussian domain, with
$\lambda > \lambda_F$, in terms of orbitals occupation numbers,
level densities and an integral involving a bivariate Gaussian in the initial
and final energies. Here we show that, using bivariate $t$-distribution,
the theory extends below from the Gaussian regime to the BW regime up to
$\lambda=\lambda_c$. This is well tested in numerical calculations for six
spinless fermions in twelve single particle states.

\end{abstract}



\maketitle
\date{}

Two-body random matrix ensembles apply in a generic way to finite interacting
many fermion systems such as nuclei \cite{Br-81,Ko-01}, atoms 
\cite{Fl-94,An-03}, quantum dots \cite{Al-01}, small metallic grains
\cite{Pa-02} etc. A common feature  of all these systems is that their
hamiltonian ($H$) consists of a mean-field one-body [$h(1)]$ plus a
complexity generating two-body [$V(2)$] interaction. With this, one has
EGOE(1+2), the embedded Gaussian orthogonal ensemble of one plus two-body
interactions operating in many particle spaces \cite{Ko-01}; for more
complete definition of EGOE(1+2) for $m$ fermions in $N$ single particle
states see \cite{EGOE}.   Most significant aspect of EGOE(1+2) is that as
$\lambda$, the strength of the random (represented by GOE) two-body
interaction [in $H=h(1) + \lambda V(2)$], changes, in terms of state density,
level fluctuations, strength functions and entropy \cite{DEN}, the ensemble
admits three  chaos markers. Firstly, it is well known that the  state
densities take Gaussian form, for large enough $m$, for all
$\lambda$ values \cite{Mo-75}. With $\lambda$ increasing, there is a chaos
marker $\lambda_c$ such that for  $\lambda \ge \lambda_c$ the level
fluctuations follow GOE, i.e. $\lambda_c$ marks the transition in the nearest
neighbor spacing distribution from Poisson to Wigner form \cite{Ja-97}.  As
$\lambda$ increases further from $\lambda_c$, the strength functions 
(for $h(1)$ basis states) change from Breit-Wigner (BW) to Gaussian form and
the transition point is denoted by  $\lambda_F$ \cite{Fl-97}. The $\lambda_c
\leq \lambda \leq \lambda_F$  region is called BW domain  and the $\lambda >
\lambda_F$  region is called Gaussian domain.  As we increase $\lambda$ much
beyond $\lambda_F$, there is a chaos marker $\lambda_t$ around which
different definitions of entropy, temperature etc. will coincide and also
strength functions in $h(1)$ and $V(2)$ basis will coincide. Thus $\lambda
\sim \lambda_t$ region is called the thermodynamic region \cite{Ks-02,Ko-03}. 

With the three chaos markers $\lambda_c$, $\lambda_F$ and $\lambda_t$, EGOE
generates statistical spectroscopy, i.e. smoothed forms for state densities,
orbit occupancies, strength sums [for example Gamow-Teller (GT) sums  in
nuclei, electric dipole ($E1$) sums in atoms], transition strengths themselves
[for  example: electric quadrupole($E2$), magnetic dipole ($M1$) and GT
strengths in nuclei, $E1$ strengths in atoms and molecules etc.], information
entropy in wavefunctions and transition strengths etc. The EGOE Gaussian
state densities are being used to generate a theory (valid for $\lambda >
\lambda_c$) for level densities with interactions \cite{Ho-03}. Similarly,
theory for orbit occupancies and strength sums, as valid in BW to Gaussian
regimes (i.e. for $\lambda > \lambda_c$) has been developed \cite{Ko-03}.
However, for transition strengths (experimentally they are most important
for probing wavefunctions structure of a quantum system), a theory valid 
only in the Gaussian domain is available \cite{Fr-88,To-86,Ks-00}. Although
a theory was given by Flambaum et al for BW domain \cite{Fl-94,Fl-96,Fl-98},
it is well known  to underestimate the exact values by a factor of 2
\cite{Fl-96,Ks-00}.  Thus, a major gap (see the discussion in \cite{Ks-00})
in understanding transition strengths is in extending the theory that works
in  the Gaussian domain, well into the BW domain. The purpose of this paper
is  to show that the bivariate $t$-distribution known in statistics will
bridge this gap. As in Refs. \cite{Fl-98,Ks-00}, we restrict ourselves to
one-body transition operators. 

Given a Hamiltonian $H$ and its $m$-particle eigenstates $ \l|E\ran$, the 
transition strengths generated by a one-body transition operator $\cal{O}$
are  denoted by $\l|\lan E_f \mid {\cal{O}} \mid E_i\ran\r|^2$; ${\cal
O}=\sum_{\alpha,\beta}\; \epsilon_{\alpha \beta} a^\dagger_\alpha
a_\beta$ where $a^\dagger_\alpha$ creates a particle in the single
particle state $\alpha$ and $a_\beta$ destroys a particle in the
state $\beta$.  Now the bivariate strength densities are defined by
\be
I^{H,m}_{biv; \co}(E_i,E_f) = \lan\lan \cod \delta (H
- E_f ) \co \delta (H-E_i) \ran\ran^{m} = \lan\lan \cod \co
\ran\ran^{m}\; \rho^{H;m}_{biv; \co}(E_i,E_f)\;.
\ee
In Eq. (1), $\lan\lan\;\;\ran\ran$ denotes trace. Note that $I_{biv;\co}$ is
square of the matrix elements of $\co$ in $H$ eigenstates weighted by the
state densities at the initial and final energies and the corresponding
$\rho_{biv;\co}$ is  normalized to unity. Moreover, one-body transition
operators $\co$ will not change $m$. The bivariate moments of
$\rho_{biv;\co}$ are defined by $M_{pq}=\lan\lan\cod H^q \co
H^p\ran\ran^m/\lan\lan\cod \co\ran\ran^m$.  With $M_{10}=\epsilon_i$ and
$M_{01}=\epsilon_f$ defining the  centroids of its two marginals,
the  bivariate central moments of $\rho_{biv;\co}$ are given by
\be
\mu_{pq}=\lan\lan \cod \;\l(H-\epsilon_f\r)^q \;
\co \; \l(H-\epsilon_i\r)^p \ran\ran^{m}\;
/\;\lan\lan\cod \co \ran\ran^{m}\;\;.
\ee
Most important of these are $\mu_{20}=\sigma_i^2$ and $\mu_{02}=\sigma_f^2$,
the variances of the two marginals and $\zeta=\mu_{11}/\sigma_i \sigma_f$, 
the bivariate correlation coefficient. 

For EGOE(1+2), going well into the Gaussian domain [then EGOE(1+2) will be 
effectively EGOE(2)], it is well established that the bivariate strength 
densities take bivariate Gaussian form (this applies to nuclei 
\cite{Fr-88,To-86}),
\be
\barr{l}
\rho_{biv; \co}(E_i,E_f) \stackrel{\lambda >> \lambda_F} {\longrightarrow}
\rho_{biv-\cg;\co}(E_i,E_f;\epsilon_i,\epsilon_f,\sigma_i,\sigma_f,\zeta) =
\dis\frac{1}{2\pi \sigma_i \sigma_f \sqrt{(1-\zeta^2)}}\;\; \times \\
\exp-\dis\frac{1}{2(1-\zeta^2)}\l\{\left(\frac{E_i - \epsilon_i}{\sigma_i}
\right)^2 - 2\zeta \left(\frac{E_i-\epsilon_i}{\sigma_i}\right)
\left(\frac{E_f-\epsilon_f}{\sigma_f}\right) + 
\left(\frac{E_f-\epsilon_f}{\sigma_f}\right)^2\;\;
\right\}\;.
\earr
\ee
An immediate question is how to extend this result well into the BW domain
and up to $\lambda_c$ (note that GOE fluctuations operate for $\lambda >
\lambda_c$ and hence in this regime it is possible to consider smoothed 
transition strengths). In a recent work, Angom et al \cite{An-04} showed
that  strength functions covering the BW to Gaussian regimes can be well
represented by Student's  $t$-distribution. Following this result, here we
conjecture that the bivariate strength density $\rho_{biv; \co}$ in Eq. (1)
can be represented by the bivariate $t$-distribution,
\be
\barr{l}	
\rho_{biv-t; \co}(E_i,E_f; \epsilon_i, \epsilon_f, \sigma_1, \sigma_2,
\zeta; \nu)=
\frac{1}{2\pi\sigma_1 \sigma_2\sqrt{1-\zeta^2}} \;\;\times \\
\l[1+ \frac{1}{\nu (1 - \zeta^2)}\left\{ \left(\frac{E_i - 
\epsilon_i}{\sigma_1}
\right)^2 - 2\zeta \left(\frac{E_i-\epsilon_i}{\sigma_1}\right)
\left(\frac{E_f-\epsilon_f}{\sigma_2}\right) + 
\left(\frac{E_f-\epsilon_f}{\sigma_2}\right)^2
\right\}\right]^{-\frac{\nu + 2}{2}}\;\;,\;\nu \geq 1\;.
\earr
\ee
Properties of $\rho_{biv-t}$ are given in \cite{Jo-72,Hu-90}. Most important
is that for $(\nu=1,\zeta=0)$, $\rho_{biv-t}$ gives  bivariate BW (called
bivariate Cauchy in statistics) distribution and as $\nu \rightarrow \infty$,
$\rho_{biv-t}$ becomes bivariate Gaussian. Thus it has the correct limiting
forms and the intermediate shapes are largely determined by the $\nu$
parameter. The marginal distributions of $\rho_{biv-t}$ are easily seen to be
univariate $t$-distributions. In Eq. (4), in general $\epsilon_i$ and
$\epsilon_f$ are the centroids of the two marginals of $\rho_{biv-t}$,
however $\sigma_1$ and $\sigma_2$ will approach the marginal widths
$\sigma_i$ and $\sigma_f$ only in the limit $\nu \rightarrow \infty$, i.e.
for the bivariate Gaussian given in Eq. (3).  In-fact, the second central
moments $\mu_{20}=\sigma_i^2$ and $\mu_{02}=\sigma_f^2$ are related to 
$\sigma^2_1$ and $\sigma^2_2$ by $\mu_{20}=\frac{\nu}{\nu-2}\;\sigma_1^2$ and
$\mu_{02}=\frac{\nu}{\nu-2}\;\sigma_2^2$ for $\nu > 2$. However $\zeta$
remains to be the bivariate correlation coefficient. Exceptions to all these
will occur for $\nu \leq 2$ and here (this happens only when $\lambda$ is
very close to $\lambda_c$) one has to use quartiles (i.e. spreading widths)
to define $\sigma_1$, $\sigma_2$ etc.; see \cite{Jo-72,Hu-90} for details. In
order to test the  applicability of the $t$-distribution, nuclear shell model
calculations are performed for isoscalar $E2$ transitions in $^{22}$Na
nucleus. Fig. 1 shows the results for $\lambda=0.4$ and 1 in the shell model
hamiltonian $H=h(1) + \lambda\,V(2)$; $\lambda=1$ gives realistic nuclear
hamiltonian. The parameters $\sigma_1$ and $\sigma_2$ in Eq. (4) are
determined via their relation to $\mu_{20}$ and $\mu_{02}$.  The value of
$\zeta=0.88$ is used as given by the exact $E2$ strengths.  Clearly (ignoring
the deviations near the ground states), the $t$-distribution gives a good
description of the transition strengths with $\nu=6$ for $\lambda=0.4$ and 
with a large $\nu$ value, as expected, for $\lambda=1$.

In larger spectroscopic spaces, instead of using a  single $t$-distribution,
to represent transition strength densities, it is more appropriate to 
partition the space. Decomposing the space into subspaces defined by $h(1)$
eigenvalues $\ce$, constructing the strength distribution generated by $h(1)$
alone, spreading this distribution by convolution with a $t$-distribution
generated by $V(2)$ and then applying some simplifying assumptions, as
described in detail in \cite{Ks-00} where this procedure is applied to
bivariate Gaussian spreadings, it is seen that the transition strengths can
be given by,
\begin{subequations}
\be
\l|\lan E_f \mid {\cal{O}} \mid E_i\ran\r|^2 = \sum_{\alpha, \beta}\;
\l|\epsilon_{\alpha \beta}\r|^2
\lan n_\beta(1-n_\alpha)\ran^{E_i} \overline{D(E_f)}\;\; {\cal F}\;;
\ee
\be 
{\cal F} = \dis\int^{+\infty}_{-\infty} \rho_{biv-t; {\cal{O}}}(E_i,\; E_f;\; 
{\cal{E}}_i,\; {\cal{E}}_f= {\cal{E}}_i- \epsilon_\beta+\epsilon_\alpha, \; 
\sigma_1, \; \sigma_2, \; \zeta; \nu)\;d{\cal{E}}_i
\ee
\end{subequations}
In Eq. (5a) $\overline{D}(E_f)$ denotes mean-spacing at the energy $E_f$,
$\epsilon_{\alpha \beta}$ are single particle matrix elements of $\co$ and
$\lan n_\beta(1-n_\alpha)\ran^{E_i} \sim \lan n_\beta\ran^{E_i}\,
\lan(1-n_\alpha)\ran^{E_i}$ with $\lan n_\alpha \ran^{E_i}$ giving
occupation probability for the single particle state or orbital $\alpha$.
Most remarkable is that  the integral for ${\cal F}$ in Eq. (5b) can be
carried out exactly for any $\nu$ and this gives,
\be
{\cal F} =
\frac{\Gamma(\frac{\nu+1}{2})}{\sqrt{\pi}\Gamma( \frac{\nu}{2})}\;\;
\frac{1} {\sqrt{\nu(\sigma_1^2 + \sigma_2^2 -2 \zeta 
\sigma_1\sigma_2)}} \;\;\left[ 1 + \frac{\Delta^2}
{\nu(\sigma_1^2 + \sigma_2^2 -2 \zeta 
\sigma_1\sigma_2)} \right]^{-\frac{\nu + 1}{2}}\;;\;\;
\Delta=E_f-E_i+\epsilon_\beta-\epsilon_\alpha
\ee
Note that, for $\nu >2$, $\sigma_1$ and $\sigma_2$ are related (as given
above) to the marginal variances $\mu_{20}$ and $\mu_{02}$ of $\l|\lan E_f
\mid {\cal{O}} \mid  E_i\ran\r|^2$. Also, the correlation coefficient $\zeta
\sim \lan \cod V \co V \ran/\lan \cod \co\ran \lan V V \ran$; see
\cite{Fr-88}. More importantly, as $\nu \rightarrow \infty$, Eq. (6) goes
exactly to  Eq. (6) of \cite{Ks-00} as it should be.  

To test the theory given by Eqs. (5a) and (6), numerical calculations are
carried out for various $\lambda$ values using 25 member EGOE(1+2) ensemble
$\{H\} = h(1) + \lambda \{V(2)\}$ in the 924 dimensional $N=12$, $m=6$ space;
$h(1)$ is defined by the single particle energies $\epsilon_i=(i)+(1/i)$,
$i=1,2,\ldots,12$ and the variance of $V(2)$ matrix elements in two-particle
spaces is chosen to be unity.  The one-body transition operator employed in
the calculations is $\co = a^\dagger_2 a_9$ as in \cite{Ks-00}. For the system
considered, $\lambda_c \sim 0.06$, $\lambda_F \sim  0.2$ and $\lambda_t \sim
0.3$. Results for six different $\lambda$ values, going from BW to Gaussian
domains, are shown in Fig. 2. Clearly Eqs. (5a) and (6) obtained via the
$t$-distribution describe  the exact EGOE(1+2) transition strengths as we go
from the BW domain with $\lambda=0.08$ to the Gaussian domain with
$\lambda=0.3$ with $\nu$ changing from 2.4 to 14;  $\nu \sim 2-6$ for
$\lambda_c < \lambda < \lambda_F$ and $\nu \sim 6-15$ for $\lambda_F < \lambda
< \lambda_t$. The exact distributions give $\zeta \simeq 0.5$ but in the fits
$\zeta$ is also varied (see Fig. 2)  and this to some extent takes into
account some of the approximations that led to the simple form given by Eqs.
(5a) and (6). More importantly, the results in Fig. 2  confirm that we have a
good method for the calculation of transition strengths in BW domain.  A
calculation is also performed for $\lambda=0.06$ by fixing $\sigma_1$ and
$\sigma_2$ using the spreading widths of the marginals of the strength
distribution and using $\zeta$ value same as that obtained for $\lambda=0.08$.
Then the deduced $\nu$ value is $1.5$. This and the comparisons in  Fig. 2
clearly emphasize the role of the bivariate correlation coefficient $\zeta$ in
BW domain and without $\zeta$  it is not possible to get a meaningful
description (it should be mentioned that the theory in the BW domain given
before \cite{Fl-98,Ks-00} uses only  the marginals of the $t$-distribution
with $\nu=1$ and $\zeta=0$). Thus all the problems seen before
\cite{Ks-00,Fl-96} in the  BW domain are cured by the bivariate-$t$
distribution with the two parameters $(\nu,\,\zeta)$. 

In conclusion, random matrix ensembles generated by a mean-field plus a
random two-body interaction generate three chaos markers. They in-turn
provide a basis for statistical  spectroscopy. The theory for transition
strengths is now extended (from Gaussian domain) to BW domain down up to the 
$\lambda_c$ marker by employing bivariate $t$-distribution. With atoms
exhibiting a clear transition from BW to Gaussian domain (an example for CeI
to SmI atoms was shown in \cite{An-04}), it is expected that the theory
given by Eqs. (5a) and (6) will be useful in the calculation of 
dipole transition strengths in the quantum chaotic domain of atoms.

\newpage

\begin{figure}
\includegraphics[width = 6in, height = 8in]{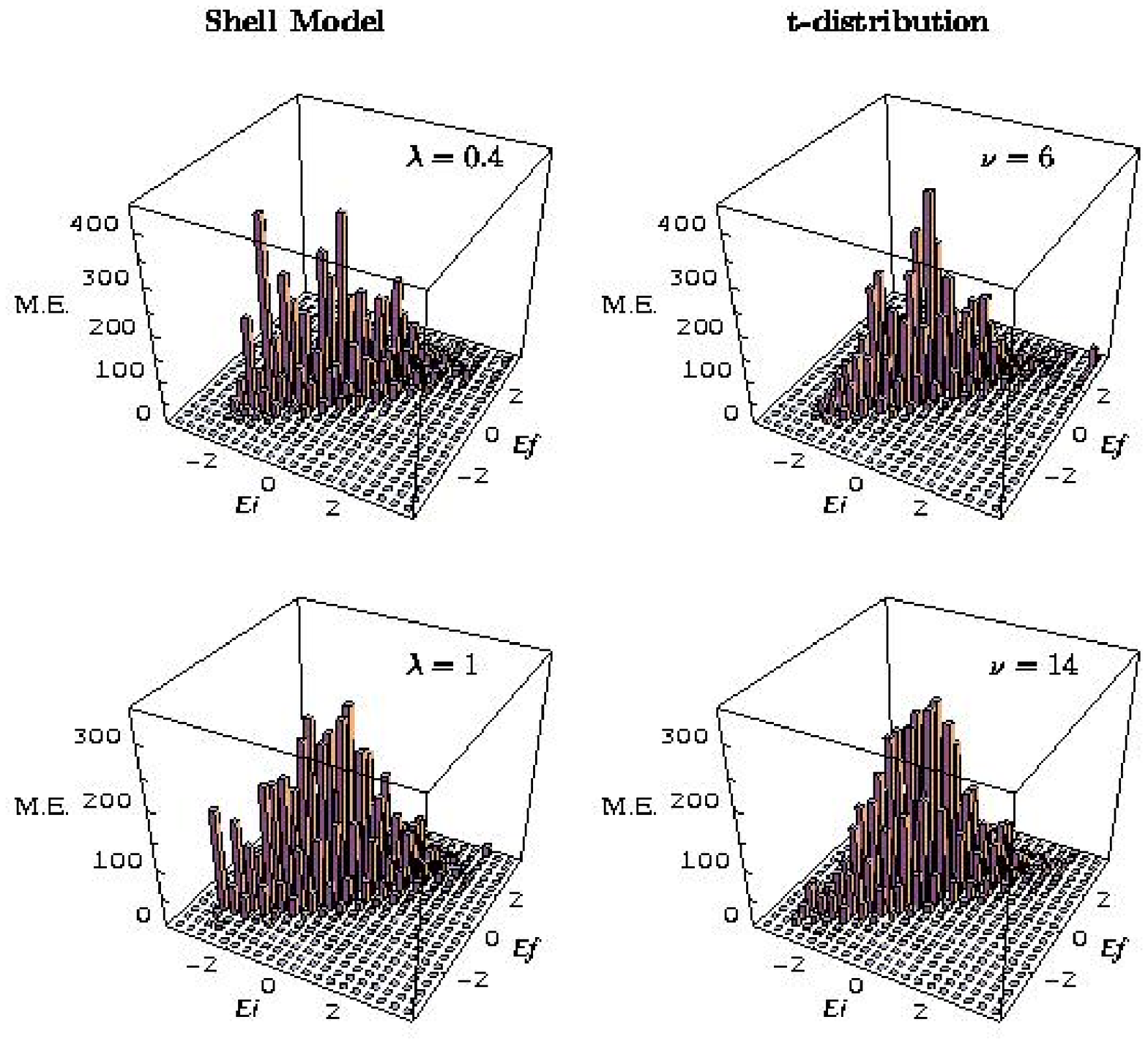}

\caption{$E2$ transition strengths vs ($Ei$, $Ef$). The $E2$ transitions
considered are $(0^+,0)$ to $(2^+,0)$; with $J$, $T$ and $\pi$ denoting
angular momentum, isospin and parity, nuclear levels are denoted by $(J^\pi,
T)$. Calculations are for  6 valence nucleons in $(2s1d)$ shell and the
hamiltonian matrix dimensions for these are 71 and 307 respectively. The
hamiltonian employed is  $H(\lambda)= h(1) + \lambda V(2)$ with the single
particle energies defining $h(1)$ and two particle matrix elements defining
$V(2)$ taken from \cite{Bw-88} and references therein (they define the so
called Wildenthal's W-interaction). The proton and neutron effective charges
for the $E2$ operator are $e_p=1.29$ and $e_n=0.49$ respectively. All the
calculations  are carried out using the  OXBASH computer  code for Windows PC
(2005-05 version) \cite{Ox-99}. In the figures the energies $Ei$ and $Ef$ 
are the energies of $(0^+,0)$ and $(2^+,0)$ levels respectively and they are
zero centered and scaled to unit width. Similarly M.E. stands for $E2$
transition strengths and they are in units of $e^2fm^4$. The vertical bars in
the figures give the total strength in a given bin area; in constructing the
histograms bin size of 0.3 is taken for both $Ei$ and $Ef$. Although Eq. (4)
is for EGOE(1+2), which is for spinless fermion systems, it can be applied
directly to the shell model with good $(J^\pi,T)$ states as described in
\cite{Go-04}.}

\label{shlmdl}
\end{figure}

\begin{figure}
\includegraphics[width = 6in, height = 8in]{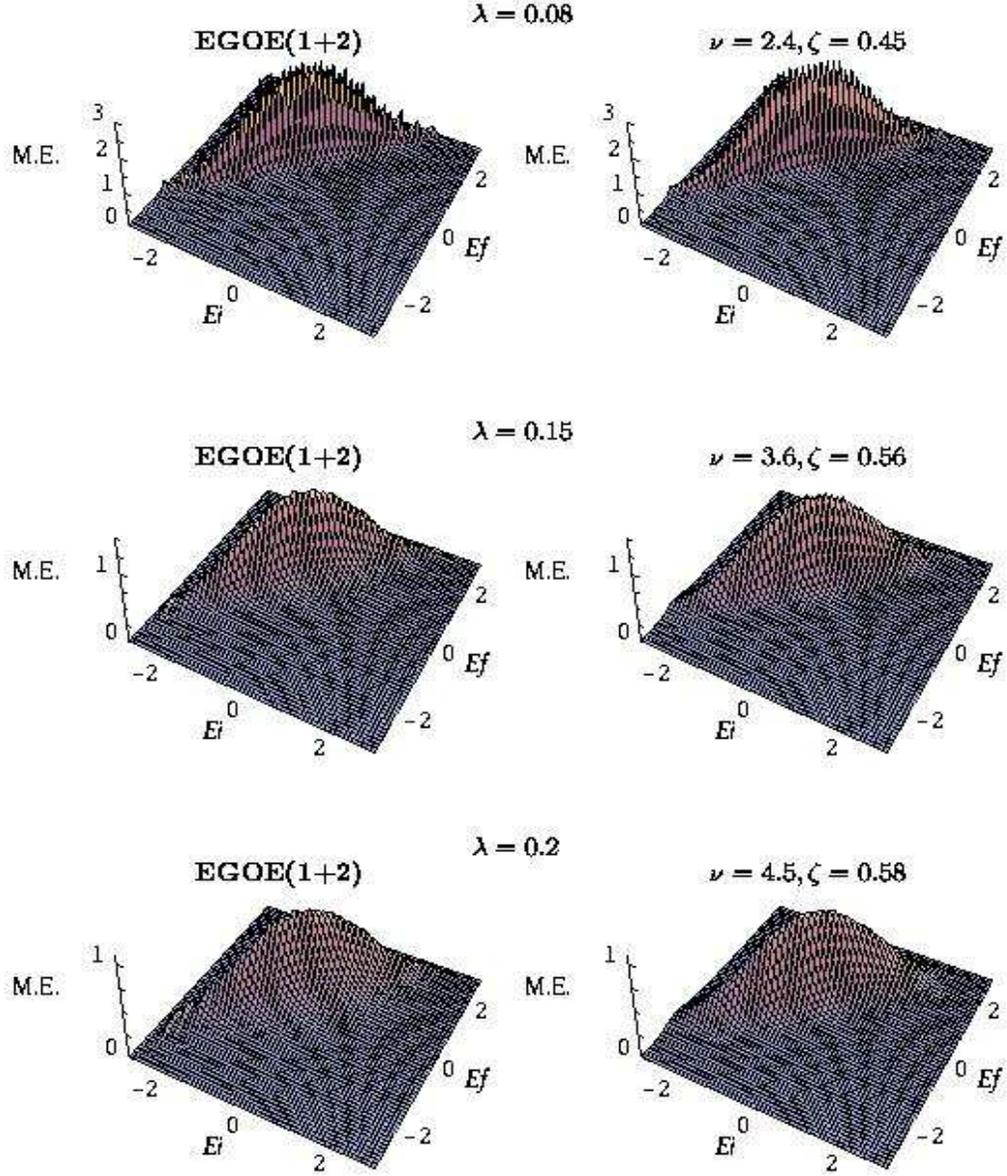}

\caption{Transition strengths $\l|\lan E_f \mid \co \mid E_i \ran\r|^2$ vs
$(Ei,Ef)$ for $\lambda=0.08$, 0.15, 0.2, 0.25, 0.28 and 0.3. In the figures 
$Ei = {\hat{E}}_i = (E_i-\epsilon)/\sigma$ and $Ef = {\hat{E}}_f =
(E_f-\epsilon)/\sigma$ where $\epsilon$ and $\sigma$ are the centroids and
widths of the state densities. Similarly M.E. stands for the strengths
$\l|\lan E_f \mid \co \mid E_i \ran\r|^2$. The EGOE(1+2) system and the
one-body transition operator $\co$ are defined in the text. In all the
calculations the strengths in the window ${\hat{E}}_i \pm
\frac{\Delta^\prime}{2}$ and ${\hat{E}}_f \pm \frac{\Delta^\prime}{2}$ are
summed and plotted at (${\hat{E}}_i$, ${\hat{E}}_f$); $\Delta^\prime$ is
chosen to be $0.1$. It should be noted that the total strength is 252. As
$\lambda$ changes from 0.08 to 0.3, the $\nu$ value changes from 2.4 to 14 and
the bivariate correlation coefficient $\zeta$ changes from 0.45 to 0.62. Note 
the change in the scales for M.E. in the figures.}   

\label{egoe}
\end{figure}   

\begin{figure}
\includegraphics[width = 6in, height = 8in]{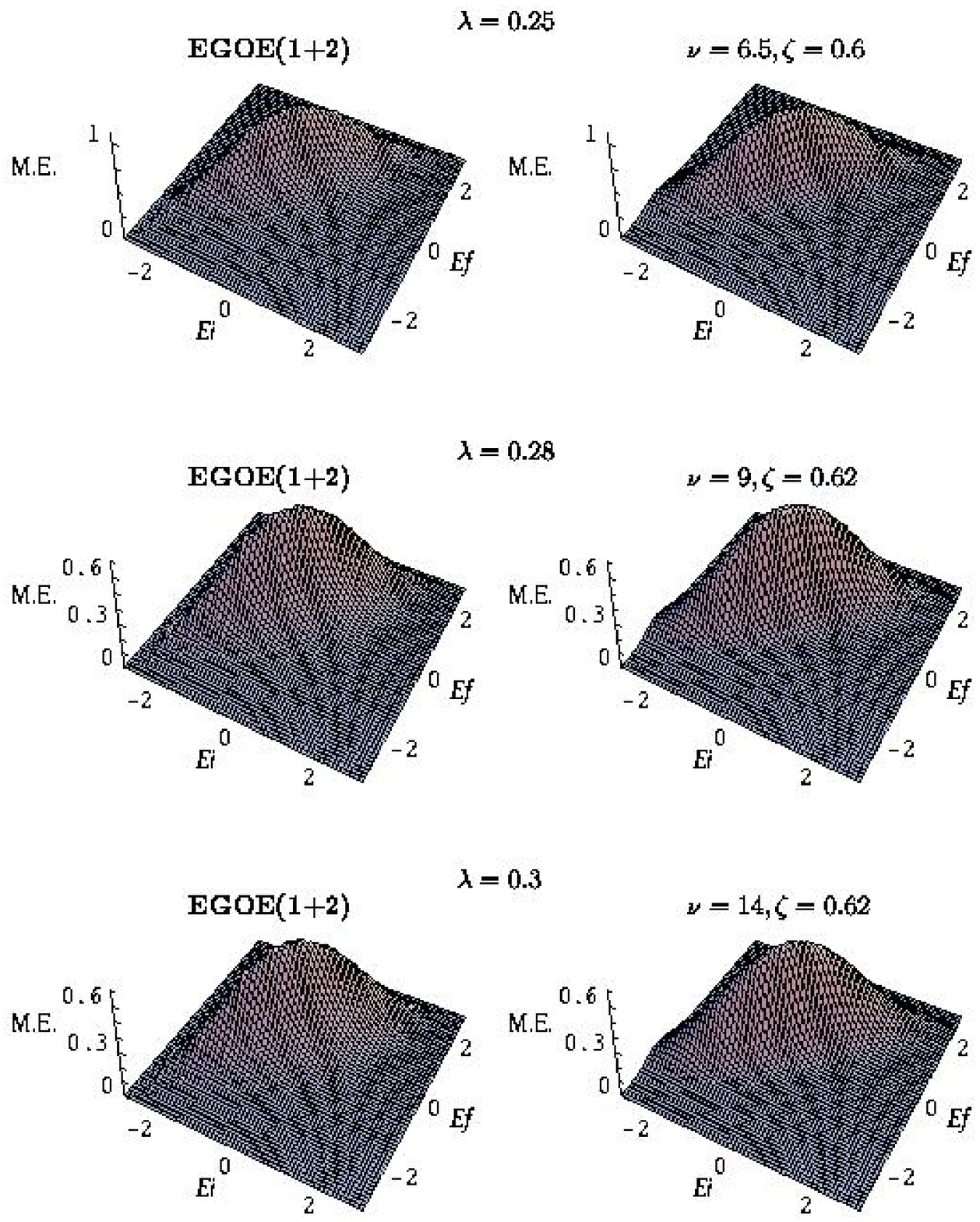}

\vskip 2cm
FIG. 2 (Cont'd)   
\end{figure}

\ed
\begin{thebibliography}{99}
\bibitem{EGOE} The EGOE(2) for $m$ ($m > 2$) spin-less
fermion systems with the particles distributed say in $N$ single particle
states ($\l| \l. \nu_\alpha \ran\r.$, $\alpha=1,2,\ldots,N$) is generated  by
defining the Hamiltonian H, which is two-body, to be GOE in the 2-particle
space and then propagating it to the $m$-particle spaces by using the
geometry (direct product structure) of the $m$-particle  spaces. Then 
EGOE(1+2) is defined by  $\{H\} = h(1) + \lambda \{V(2)\}$  where $\{\;\}$
denotes an ensemble. The mean-field one-body Hamiltonian $h(1)=\sum_\alpha
\epsilon_\alpha n_\alpha$ is a fixed one-body operator defined by the single
particle energies $\epsilon_\alpha$ with average spacing $\Delta$ (note that
$n_\alpha$ is the number operator for the single particle state $\l. \l|
\alpha\r. \ran$). The $\{V(2)\}$ is EGOE(2) with  unit variance for the
two-body matrix elements and $\lambda$ is the strength of the two-body 
interaction (in units of $\Delta$). Thus, EGOE(1+2) is defined by the four 
parameters $(m,N,\Delta,\lambda)$ and without loss of generality we put
$\Delta=1$.
\end{thebibliography}
